\documentclass{article}
\usepackage{epsfig}
\usepackage{cite}
\usepackage{float}
\usepackage{mathrsfs}
\usepackage{amsfonts}
\usepackage{array}
\usepackage{epsfig}
\usepackage{amsmath}    
\usepackage{amssymb}   
\usepackage{graphicx}   
\usepackage{verbatim}   
\usepackage{color}      
\usepackage{subfigure}  
\usepackage{subfiles}
\usepackage{slashed}

\def\beq{\begin{equation}}
\def\eeq{\end{equation}}
\def\beqa{\begin{eqnarray}}
\def\eeqa{\end{eqnarray}}
\def\ba{\begin{eqnarray}}
\def\ea{\end{eqnarray}}
 
\newlength{\dinwidth} \newlength{\dinmargin}
\title{ }
\author{Alon E Faraggi }
\date{May 2023}

\begin{document}

\begin{center}
{\Large \bf String Derived Z$^\prime$ Model at an\\ \medskip
Upgraded Superconducting Super Collider}
\end{center}
\vspace{2mm}
\begin{center}
{\large Alon E. Faraggi$^{(a)}$, Marco Guzzi$^{(b)}$, Andrew McEntaggart$^{(b,c)}$}\\
\vspace{2mm}
\it{$^{(a)}$Department of Mathematical Sciences,
University of Liverpool,\\
Liverpool L69 7ZL, UK}

{\it $^{(b)}$Department of Physics, Kennesaw State University,\\
Kennesaw, GA 30144, USA}

\it{$^{(c)}$Department of Physics, Georgia Institute of Technology, \\Atlanta, GA 30332, USA}

\end{center}

\begin{abstract}

The future of collider physics is under investigation. With the High Luminosity LHC program lasting until the late 2030s, 
the next machine in the energy frontier is envisioned to appear in 30--40 years, which may be too far into 
the future to sustain the field. In this paper we explore the physics potential of an Upgraded Superconducting 
Super Collider (USSC). The Original Superconducting Super Collider (OSSC) was planned to operate at 20TeV beam 
energy, and with improved magnet technology and/or longer tunnel, one may envision that it can be extended to 
25--30TeV beam energy. Given that the decision on the OSSC construction took place in Autumn 1988 and it was planned
to start operation in the 1996-1999 period, an USSC can be constructed 10--15 years from decision and fill 
the gap between the end of HL--LHC and the future envisioned machines. 
While the main mission of the USSC will be to test the 
Standard Model and its electroweak and strongly interacting sectors, 
as a specific example we illustrate the invariant mass distribution at NNLO in QCD for a 5 TeV $Z^\prime$
in the string derived $Z^\prime$ model.

\end{abstract}

\vfill\eject

\section{Introduction}

Fundamental particle physics is at a crossroad. On the one hand the basic understanding of the constituent matter
and interactions crystallised in the Standard Particle Model that correctly accounts for all experimental observations
to date. The latest victory on this triumphant march, culminating over one hundred years of experimental research and 
discovery, has been the experimental discovery of the Higgs boson at $125.11{\rm GeV}$. The Standard Model utilises the 
theoretical framework of point Quantum Field Theories and is also used in the Standard Cosmological Model that correctly
predicts some of the observed features such as the primordial abundance of light elements. It is further noted that given 
our current understanding of the fundamental matter and interactions, it is plausible that the Standard Model 
remains unperturbed up energy scales much above the energy scales being probed in contemporary experiments, {\it i.e.}
the Grand Unified Theory (GUT) scale, at which the three gauge interactions of the Standard Model are merged into one, or 
the Gravitational Unified Theory (GUT) at the Planck scale, at which the gravitational interaction is synthesised with the gauge interactions.
On the other hand, the Standard Model leaves much to be desired. Our basic understanding of the content and composition of 
the matter in the universe is lacking. Most of this matter is hidden from us, and we do not understand why that which is
not hidden exists at all. Furthermore, most of the energy stored in the universe is not in the form of matter at all. 
Most importantly, however, is the nature of the Higgs boson and the electroweak symmetry breaking mechanism. 
We do not understand how the Standard Model can remain viable up to much higher energy scales, while maintaining the 
stability and perturbativity of its scalar sector. That is the most urgent contemporary question that requires 
further experimental elucidation. 

Much of this celebrated success was made possible by the development of collider technology. 
It is clear that collider tools will cease to provide a viable experimental probe 
once the required energy regimes will enter the Grand Desert Scenario envisioned in GUTs. 
Nevertheless, the nature of the Higgs boson and the 
mystery of electroweak symmetry breaking can only be unravelled effectively by utilising
colliding machines. The pivotal question that occupies the community at present is which 
collider facility is best at present to advance the objectives of the field. 
In this note we propose that the best option in terms of feasibility and timeline is
an Upgrade of the Super Conducting Super Collider (USSC) that was designed and partially
constructed in the early 1990s. We give some superficial arguments in favour of this
possibility that to our knowledge has not been discussed seriously so far among the 
options that are being debated. To examplify the potential utility of the USSC
we analyse the physics of a superstring derived $Z^\prime$ model in the energy 
scales that will be probed in the USSC. Additionally, we discuss briefly
other research aspects of the experiment

The lightness of the Standard Model scalar sector as compared to the 
large scales envisioned in GUTs is best accommodated in its supersymmetric
extensions, which mandate the existence of at least two electroweak Higgs 
doublets in vector--like representations. This introduces a new version 
of the hierarchy problem which is known as the $\mu$--problem. 
This problem is particularly acute in string--derived models that
reproduce the phenomenological features of the Minimal Supersymmetric
Standard Models. Namely, what prevents the $\mu$--parameter from being
of the order of the Planck scale, rather than of the order of the TeV or 
weak scales. We proposed that a solution to this problem can be found 
if the Higgs multiplets are chiral under an additional $U(1)$ symmetry
that remains unbroken down to low scales. In this case, the $\mu$--parameter
is only generated by the Vacuum Expectation Value that breaks the 
additional $U(1)$ symmetry. 

In this paper, we show the invariant mass distribution for a 5 TeV string derived $Z'$, produced in the Drell-Yan (DY) process at the USSC with $\sqrt{S}=50$ TeV of center of mass energy and compare this distribution to that at obtained at the LHC with $\sqrt{S}=13$ TeV. 

The study of gauge kinetic-mixing interactions $\chi F^{\mu\nu}F'_{\mu\nu}$ in the lagrangian~\cite{Holdom:1985ag} will be shown in a separate work.
The effect of such a renormalizable operator is important in this context because it may be produced by physics 
at energy scales above the $Z'$ breaking scale, with no suppression by the large-mass 
scale~\cite{Polchinski:1982an,Dienes:1996zr} and can affect the couplings of $Z'$s at the TeV 
scale~\cite{Rizzo:1998ut}. The interplay between the gauge kinetic--mixing parameter and gauge coupling of the $Z'$, 
as well as other $Z'$ properties, can be probed by using forward--backward asymmetry ($A_{FB}$) distributions in 
Drell--Yang scattering, where this observable can have an advantage in extra--resonance searches and as a model discriminator 
tool~\cite{London:1986dk,Rosner:1986cv,Hewett:1988xc,Cvetic:1995zs,Rosner:1995ft,Dittmar:1996my,Bodek:2001xk,
CDF:1997wdd,Accomando:2015cfa,Fiaschi:2022wgl,Ball:2022qtp,McEntaggart:2022hey}.

\section{Feasibility and timeliness}

At the outset we should attest that we are not accelerator physicists. The remarks made
here are not based on expert knowledge in accelerator physics. Nevertheless, they follow
pure logic and familiarity of historical development of collider based research over the
past 35+ years. The future of particle accelerators has undergone detailed studies by the 
American and European communities over the past few year that are published and are 
accessible online \cite{Butler:2023glv, Narain:2022qud, 
Gourlay:2022odf, EuropeanStrategyGroup:2020pow}. 
A leading experimental facility at the Energy Frontier (EF), the High--Luminosity
Large Hadron Collider (HL--LHC) will operate at CERN until the mid-- to late--2030s.
The experimental program beyond HL--LHC at CERN is currently under study and will
entail a well laid out program of Future Circular Collider (FCC) program with 
$e^+e^-$ and $hh$ phases. There isn't enough praise in the world that can be 
bestowed on CERN and its experimental program since its establishment in the 
1950s. CERN stands as the model for international collaboration and cooporation 
at the forefront of human curiosity and exploration. CERN has its experimental
program on the Energy Frontier well laid out beyond 2050. The physics 
case and the experimental instruments to explore it are planned and justified. 
 There is not much to add. 
The other contemporary global players with potential capacity in the Energy Frontier
are China and the United States. Detailed community studies are on--going
that can be found in the published summaries, with the most relevant 
being the Accelerator Frontier \cite{Gourlay:2022odf} and the Energy Frontier
\cite{Narain:2022qud}. 

These reports contain many excellent proposals that can be divided into two main
categories. Precision lepton colliders that will provide instruments 
to perform detailed studies of the Higgs particle and its properties, 
as well as precision studies at the $t{\bar t}$ threshold. The initial 
phase of this program will be an initial run at 250GeV Centre of Mass (CoM) energy 
with increasing energy up to 1TeV and the few TeV CoM regime. The colliders in this 
category include the International Linear Collider (ILC) that has been in discussion since 
the early 90s, the Compact Linear Collider (CLIC); the Circular Electron Positron Collider
(CEPC); the Cool Copper Collider (${\rm C}^3$); and High-Energy-LEptoN (HELEN) collider. 
Another class in this category is the Muon Collider that will provide a 
precision machine with reach into the multi--TeV energy scale. The 
second category are the hadron--hadron colliders that include the FCC--hh with 
100TeV CoM energy, a hadron--hadron phase in the Future Circular Collider (FCC) at CERN that will
have an initial $e^+e^-$ program, similar to the LEP/LHC program, and the 
Super Proton-Proton Collider (SPPC) with a planned 100TeV CoM energy
to operate in China after the CEPC and using the same tunnel complex
and infra--structure in a multi--staged approach, similar to that planned 
for the CERN FCC. Another category of proposed colliders are the electron-proton
colliders, {\it e.g.} the LHeC, and the electron--ion collider {\it e.g.} the EIC,
that provide more limited scope in terms of 
Higgs precision studies and/or discovery potential. 

In terms of the timeline of the current proposals, the HL--LHC will run until 
the mid--to--late 2030s. It is then envisioned that within a 10 year period 
the FCC program can be set in motion and start operation toward the mid to late 
2040s, with an initial $e^+e^-$ phase, followed by hadron--hadron phase. 
A similar timeline is envisioned for the CEPC to be followed by the SPPC. 
The ILC program may be operational by mid--to--late 2030s. The initial 
phases of the lepton colliders will run at 0.25TeV and will be limited
to precision Higgs studies. It is noted that $e^+e^-$ machines require 
relatively short accelerator R\&D, whereas the envisioned hadron machines
at 100TeV require substantial accelerator R\&D. 
The Muon Collider
requires substantial accelerator R\&D before it can be considered as 
a physics exploratory machine. The programs for the hadron machines 
envisions discovery machines at 100TeV at timeline reaching beyond the 
2050s. 

An alternative route for an hadron--hadron machine is to consider an
Upgrade of the Super Conductor Super Collider (USSC) that was being built 
in the early 1990s and was terminated by the US congress in October 1993. 
In terms of magnet technology the SSC magnets were supposed to operate 
at 6T with an initial aperture of 4cm bore that was increased to 5cm 
bore, with increased cost of the total project from 6B to 11B USD. It 
was designed at 20TeV per beam with 40TeV CoM energy and $10^{33}$/(cm$^2\cdot$s) 
total luminosity. We can consider that with modern magnet technology 
an increase of the magnetic field to 8-10T will enable the construction 
of a machine with 25--30TeV beam energy with 50--60 CoM energy. The main
advantage of this option is that the required accelerator R\&D is substantially
shortened. The project will utilise the blueprint of the SSC design, albeit with
some modification for the upgrade. The Original SSC (OSSC) site was chosen in 
October 1988 and it was initially planned to start operations in 1996, which was 
delayed to 1999. Given this timelines, the timeline for the USSC is 10--15 years
from the decision point, {\it i.e.} mid--to--late 2030s. 

In terms of the physics reach, it is clear that the main focus of experimental 
particle physics should be the study of the properties of the Higgs boson and 
its many couplings. Particularly vital is to elucidate the electroweak symmetry
breaking mechanism are the self--couplings in the Higgs potential. The optimal
machines for this purpose are the lepton machines. In that respect, however, the 
$e^+e^-$ machines at the lower energy scale are limited scope machines and
their prospects as discovery machines are limited. Hadronic machines offer a less
clean environment to study the Higgs parameters, but offer a more extensive 
energy reach and discovery potential. We should comment, however, that there 
is an erroneous cultural perception among some researchers that an accelerator
project is successful only if some new physical phenomena is discovered by it. 
We advocate that an accelerator project at the energy frontier should be 
deemed successful if it delivers on the specified properties
on which it was designed, {\it e.g.} in terms of energy and luminosity. It should 
then be considered as substantially contributing to general knowledge if it 
is able to reduce the error bars on the Standard Model parameterisation of 
the experimental observables. 

In terms of location, the USSC could be built in the US, China and/or Europe. 
The technical know--how exist in Europe and the US and will need to be acquired 
in China. All three have the industrial capacity to carry out the project. 
It is the prevailing culture in this field that projects of this scale are 
global projects with active collaborations from all over the globe and 
from different societal regimes. 
However, all three are conducting community studies that are well underway. 
Borrowing the cost estimates
from the Snowmass studies \cite{Gourlay:2022odf}, we may estimate the 
cost of the project between 10B-20B USD. The question then is which 
entities possess the financial resources to fund the project. Given 
that the required accelerator technology is off the shelf technology
Saudi Arabia (SA) can also fund the construction of the USSC 
as a Middle East (ME) project, at a cost equivalent to the establishment 
of a few football teams. Given that SA will need to buy the technical 
know--how and to build the industrial capacity, we can multiply the 
projected cost by a factor of five, {\it i.e.} 50B--100B USD. 
As a ME project the USSC can build on the experience gained in the SESAME
ME project and subject to geological studies base the USSC at the SESAME 
site. 
%
%
The construction of the USSC and successful delivery on the specified 
design parameters would be by itself an enormous success. 
Following the CERN experience of providing a platform for cooperation, the USSC can serve as a
Project for Peace, promoting curiosity driven collaboration. 

\section{The Physics Case}

In this section we will briefly discuss in general the physics case for the USSC. 
A more detailed analysis will be presented in a forthcoming publication, including 
a detailed technical proposal and elaboration of the physics case. We comment that 
the physics case of a project of this financial commitment must be based on 
improving the measurement of the parameters of the prevailing theory, {\it i.e.}
measurement of the Standard Model parameters. The discovery of new physics can 
only be regarded as an added bonus. On the other hand, there is a non--zero chance
that the USSC will stumble on the new physics threshold that is anticipated to exist
in connection with the electroweak symmetry breaking mechanism. While this cannot be guaranteed,
there is potentially a big prize to be won. As an example, we analyse 
the invariant mass distribution an NNLO in QCD for a 5 TeV $Z^\prime$ produced at 
in pp collisions at $\sqrt{S}=13~{\rm and}~\sqrt{50}$ TeV. 
It is clear that more elaborate study is required for the general physics case. 
We note, however, that proposals for projects with intermediate energy scales 
between the current LHC energy scale at 14TeV and the projected FCC--hh energy 
scale at 100 TeV are being discussed at the order of double the LHC energy, 
{\it i.e.} 28 TeV CoM energy. The USSC on the other hand will operate at 50-60 TeV 
CoM energy. The USSC therefore targets the mid--energy range between the LHC and 
the planned FCC--hh. It is clear that the physics case has to be based on improving the 
measurements of the Standard Model parameters and at present the primary 
interest is the measurements of the couplings of the Higgs field, {\it i.e.}
the coupling of the Higgs field to the Standard Model fermions and vector fields, 
and most importantly the Higgs self--couplings that will elucidate the nature 
of the Higgs potential and the electroweak symmetry breaking mechanism. 
It is further evident that the optimal machine to preform these measurements 
is a lepton accelerator. However, the scope of a $e^+e^-$ accelerator at 0.25--0.5 TeV
will be limited to Higgs studies. On the other hand the USSC will have by far 
greater chance of stumbling on the much anticipated layer of new physics which 
is associated with the electroweak symmetry breaking mechanism. While this is 
not guaranteed, the mere construction of the ME USSC will be regarded as a 
transformative contribution for regional development. 

Details of the string derived $Z^\prime$ model can be found in the literature 
\cite{Faraggi:2014ica, DelleRose:2017vvz, Faraggi:2022emm, McEntaggart:2022hey}. 
%
%
%
%
String inspired $Z^\prime$ models have been of interest since the mid--eighties
(see {\it e.g.} \cite{Zwirner:1987kxa,Hewett:1988xc, Leike:1998wr, King:2020ldn})
due to the appearance of Grand Unified Theoretical (GUT) structures in heterotic--string
compactifications. However, the construction of string derived models that 
allow for an extra $U(1)$ symmetry to remain unbroken down to low scales
proved to be challenging. The reason is that the symmetry breaking pattern 
$E_6\rightarrow SO(10)\times U(1)_A$ in string constructions results in an 
anomalous $U(1)_A$ that cannot remain unbroken down to lower energy scales
\cite{Cleaver:1997rk}. 
Viability of a low scale $Z^\prime$ vector boson mandates the construction 
of string models with anomaly free $U(1)_A$. This can be achieved by 
enhancing $U(1)_A$ to a non--Abelian symmetry as in ref. \cite{Bernard:2012vf},
or by utilising self--duality under the Spinor--Vector Duality (SVD) 
\cite{Faraggi:2006pk,Angelantonj:2010zj, Faraggi:2011aw},
as in ref. \cite{Faraggi:2014ica}. 
We further note that the string derived $Z^\prime$ models 
\cite{Faraggi:2014ica} has an extended Pati--Salam symmetry at the string scale. 
The heavy Higgs states in the string model that are available to break the 
extended Pati--Salam symmetry imply that the breaking scale of this linear
combination coincides with the breaking scale of the Pati--Salam symmetry. 
In the string--derived $Z^\prime$ model the observable and hidden sector
gauge symmetries are given by: 
\begin{eqnarray}
{\rm observable} ~: &~~SO(6)\times SO(4) \times
U(1)_1 \times U(1)_2\times U(1)_3 \nonumber\\
{\rm hidden}     ~: &SO(4)^2\times SO(8)~~~~~~~~~~~~~~~~~~~~~~~~~~~~~~~\nonumber
\end{eqnarray}
The string spectrum contains the required massless states
to break the GUT symmetry to the Standard Model.
The two $U(1)$ symmetries $U(1)_1$ and $U(1)_2$ are anomalous in the string 
model
\begin{equation}
{\rm Tr}U(1)_1= 36 ~~~~~~~{\rm and}~~~~~~~{\rm Tr}U(1)_3= -36.
\label{u1u3}
\end{equation}
and the $E_6$ combination, 
\begin{equation}
U(1)_\zeta ~=~ U(1)_1+U(1)_2+U(1)_3~,
\label{uzeta}
\end{equation}
is anomaly free
and can be part of an unbroken $U(1)_{Z^\prime}$ symmetry at low scales.
The Pati--Salam gauge symmetry is reduced by the VEVs of the heavy 
Higgs states ${\cal H}$ and
$\overline{\cal H}$, with charges under the
Standard Model gauge group given by:
\begin{align}
\overline{\cal H}({\bf\bar4},{\bf1},{\bf2})& \rightarrow u^c_H\left({\bf\bar3},
{\bf1},~~\frac 23\right)+d^c_H\left({\bf\bar 3},{\bf1},-\frac 13\right)+{\overline {\cal N}}\left({\bf1},{\bf1},0\right)+
                             e^c_H\left({\bf1},{\bf1},-1\right)
                             \nonumber \\
{\cal H}\left({\bf4},{\bf1},{\bf2}\right) &
\rightarrow  u_H\left({\bf3},{\bf1},-\frac 23\right)+
d_H\left({\bf3},{\bf1},~~\frac 13\right)+  {\cal N}\left({\bf1},{\bf1},0\right)+
                            e_H\left({\bf1},{\bf1},~~1\right)\nonumber
\end{align}
The VEVs along the ${\cal N}$ and $\overline{\cal N}$ flat directions leave
$N=1$ spacetime supersymmetry and the $U(1)$ combination
\begin{equation}
U(1)_{{Z}^\prime} ~=~
\frac{1}{5} (U(1)_C - U(1)_L) - U(1)_\zeta
~\notin~ SO(10),
\label{uzpwuzeta}
\end{equation}
unbroken below the string scale. This $U(1)_{Z^\prime}$ symmetry is anomaly free provided that $U(1)_\zeta$ is anomaly free.

\begin{table}[!ht]
\noindent
{\small
\begin{center}
{
\begin{tabular}{|l|cc|c|c|c|}
\hline
Field &$\hphantom{\times}SU(3)_C$&$\times SU(2)_L $
&${U(1)}_{Y}$&${U(1)}_{Z^\prime}$  \\
\hline
$\hat{Q}_L^i$&    $3$       &  $2$ &  $+\frac{1}{6}$   & $-\frac{2}{5}$   ~~  \\
$\hat{u}_L^i$&    ${\bar3}$ &  $1$ &  $-\frac{2}{3}$   & $-\frac{2}{5}$   ~~  \\
$\hat{d}_L^i$&    ${\bar3}$ &  $1$ &  $+\frac{1}{3}$   & $-\frac{4}{5}$  ~~  \\
$\hat{e}_L^i$&    $1$       &  $1$ &  $+1          $   & $-\frac{2}{5}$  ~~  \\
$\hat{L}_L^i$&    $1$       &  $2$ &  $-\frac{1}{2}$   & $-\frac{4}{5}$  ~~  \\
\hline
$\hat{D}^i$       & $3$     & $1$ & $-\frac{1}{3}$     & $+\frac{4}{5}$  ~~    \\
$\hat{{\bar D}}^i$& ${\bar3}$ & $1$ &  $+\frac{1}{3}$  &   $+\frac{6}{5}$  ~~    \\
$\hat{H}_{\textrm{vl}}^i$       & $1$       & $2$ &  $-\frac{1}{2}$   &  $+\frac{6}{5}$ ~~    \\
$\hat{{\bar H}}_{\textrm{vl}}^i$& $1$       & $2$ &  $+\frac{1}{2}$   &   $+\frac{4}{5}$   ~~  \\
\hline
$\hat{S}^i$       & $1$       & $1$ &  ~~$0$  &  $-2$       ~~   \\
\hline
$\hat{H}_1$         & $1$       & $2$ &  $-\frac{1}{2}$  &  $-\frac{4}{5}$  ~~    \\
$\hat{H}_2$  & $1$       & $2$ &  $+\frac{1}{2}$  &  $+\frac{4}{5}$  ~~    \\
\hline
$\hat{\phi}$       & $1$       & $1$ &  ~~$0$         & $-1$     ~~   \\
$\hat{\bar\phi}$       & $1$       & $1$ &  ~~$0$     & $+1$     ~~   \\
\hline
\hline
$\hat{\zeta}^i$       & $1$       & $1$ &  ~~$0$  &  ~~$0$       ~~   \\
\hline
\end{tabular}}
\end{center}
}
\caption{\label{table27rot}
Supermultiplet spectrum and
$SU(3)_C\times SU(2)_L\times U(1)_{Y}\times U(1)_{{Z}^\prime}$
quantum numbers, with $i=1,2,3$ for the three light
generations. The charges are displayed in the
normalisation used in free fermionic
heterotic--string models. }
\end{table}

Anomaly cancellation down to low scale 
mandates the existence of additional chiral states in the spectrum that are 
vector--like under the Standard Model gauge group. The spectrum below the 
Pati--Salam symmetry breaking scale is summarised in table \ref{table27rot}. 
We note therefore that these extra electroweak Higgs doublets can only receive
mass when the $U(1)_{Z^\prime}$ symmetry is broken, and relates the electroweak 
symmetry breaking scale to the $U(1)_{Z^\prime}$ breaking scale. An additional
$SU(2)_L$ doublet pair is added to facilitate gauge coupling unification at the 
string scale. However, one can envision that the question of gauge coupling unification
can be addressed in other ways and that the chirality of the electroweak Higgs doublets 
under $U(1)_{Z^\prime}$ is the origin of the electroweak symmetry breaking scale. 
We then anticipate a rich new sector to appear in association with the $U(1)_{Z^\prime}$ symmetry
breaking scale.

\begin{figure} 
\centering
\includegraphics[width=0.65\textwidth]{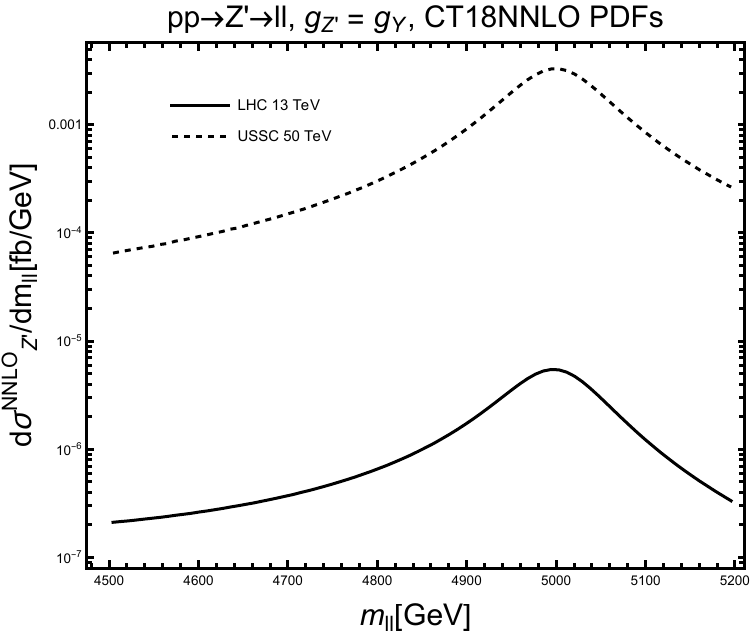}
\caption{Invariant mass distribution at NNLO in QCD for a 5 TeV $Z'$ produced in pp collisions at $\sqrt{S}=13$ TeV and $\sqrt{S}=50$ TeV.}
\label{USSC-Zp-invmass}
\end{figure}

A detailed analysis of precision studies of this string derived $Z^\prime$ model recently appeared~\cite{McEntaggart:2022hey} and
we refer the reader to this paper for the details of the methodology. In Fig.~\ref{USSC-Zp-invmass}, we illustrate the central theory predictions at NNLO in QCD for the invariant mass distribution of a 5 TeV $Z'$ produced in pp collisions at $\sqrt{S}=$ 13 TeV and $\sqrt{S}=$ 50 TeV respectively, obtained with the \texttt{MCFM-10.2.2} computer program~\cite{Campbell:2019dru} with CT18NNLO parton distribution functions (PDFs) of the proton~\cite{Hou:2019efy}. We note that the invariant mass distribution scales up by approximately 3 orders of magnitude as the center of mass energy goes from 13 TeV to 50 TeV, making the USSC a machine with novel and superior discovery potential.

\section{Conclusions}

In this article we proposed the USSC as a feasible and timely collider experiment to explore
the physics of the electroweak symmetry breaking mechanism. While the optimal machine to 
measure the Higgs parameters and properties most precisely is an $e^+e^-$ collider, the USSC
is more likely to stumble on the new physics that is anticipated to exist in connection with
electroweak symmetry breaking mechanism. Different excellent proposals are being studied
for the future collider physics program and each will have its advantages and disandvantages. 
A key advantage of the USSC proposal is that it does not require substantial accelerator R\&D
and proposes to use established accelerator technology, albeit with upgraded magnets and 
beam energy as compared to the OSSC. 
%
%
We further analysed the production 
of the string derived $Z^\prime$ at the USSC as compared to the LHC. From our analysis we note the invariant mass 
distribution scales up by approximately 3 orders of magnitude when proton beams collide at 50 TeV as compared to the LHC. 
This increases the discovery potential of the USSC.   

\section*{Acknowledgements}

The work of MG and AM is supported by the National Science Foundation under Grant no.
2112025. AEF would like to thank the CERN theory group for hospitality and discussions
and acknowledges the support of a CERN Associateship. 
MG would like to thank the Erwin Schr\"odinger International Institute for Mathematics and Physics (ESI) in Vienna, for hospitality and discussions, and partial support. 

\bibliographystyle{JHEP}
\bibliography{references}

\end{document}